\documentclass[12pt]{article}
\usepackage{latexsym,amsmath,amssymb,amsbsy,graphicx}
\usepackage[cp1251]{inputenc}
\usepackage[T2A]{fontenc}
\usepackage[russian,english]{babel}
\usepackage[space]{cite} 

\begin{document}
\bigskip

{\bf Large flares (M1 -- X7) in solar cycle 24 }

\bigskip

\centerline {E.A. Bruevich $^{a}$, T.V. Kazachevskaya $^{b}$, G.V. Yakunina $^{a}$}

\centerline {\it {$^a${ Sternberg Astronomical Institute, Lomonosov Moscow State
 University,}}}\
\centerline {\it Universitetsky pr., 13, Moscow 119234, Russia}\

\centerline {\it {$^b${ Fedorov Institute of Applied Geophysics}}}
\centerline {\it Rostokinskaya, 9, Moscow, 129128, Russia}\

\centerline {\it e-mail:  {red-field@yandex.ru, kazachevskaya@mail.ru,
yakunina@sai.msu.ru} }\

\bigskip

{\bf ABSTRACT}

The large (X-ray class > M1) and very large (X-ray class >X1) flares
(according to the observations of GOES-15 and Preliminary data from
Current Catalog of Flare Events) in solar activity cycle 24 were
analyzed. The monthly average values of optical Flare Index for 2010
-- 2016 were calculated. The values of the total energy of the flare
E $(J \cdot m^{-2})$ in the 0.1 -- 0.8 nm range at the level of the
earth's atmosphere were estimated. The energy spectrum (the
dependence of the number of flares with the full energy E from the
value of this full energy) for 115 flares of M5 -- X7 classes was
built. The comparative study of monthly averaged values of several
indices of solar activity in current cycle 24: the relative sunspot
numbers (SSN), the 10.7 cm radio flux ($F_{10.7}$), the radiation
flux in the Lyman-alpha line ($F_{L \alpha}$), the solar constant
(TSI) and the Flare Index (FI) was made.


\bigskip

\vskip12pt
\section{Introduction}
\vskip12pt

Cycle 24 has already proved to be quite different from resent
cycles. Solar cycle 24 has started on December 2008, but according
to the observations of global solar indices there was the minimal
activity until early 2010. The previous sunspot archive data since
accurate records began in 1850 show that solar cycle 24 is one of
the lowest during the cycles with recorded sunspot activity. It was
shown that there is the opposite long-scale trends during solar
cycles 22 -- 24, [Bruevich and Yakunina, 2015; Khlystov and Somov, 2012; Khlystov, 2014].

[Nagovitsyn et al.,
2012; Pevtsov et al., 2014] have analyzed the recent SSN
observations from [Pevtsov et al., 2011]. It was shown that in the
recent times the number of large sunspots was decreased, but the
number of small sunspots was increased. This change can be explained
by the gradual decline in average sunspot field strength in
accordance with the observation data. At the end of the XX-th and in
the beginning of the XXI-st centuries we can see the minimum similar
to the minimum of Dalton (the period of low SSN corresponding to low
solar activity, lasting from about 1790 to 1830) and this current
minimum is possibly a result of the superimpose of the 11-yr minimum
on the minima of the 50-yr and the 100-yr cycles. SSN and $F_{10.7}$
as the good indicators of solar activity show the reduction of their
maximum values in cycle 24 equal to approximately 30 -- 40\%.

Flare activity in solar cycles 23 and 24 also differ significantly
from previous cycles 21 and 22, this can be seen from Figure 1.
Flares appear in areas with a complex structure of the magnetic
field, which plays a determining role in flare's occurrence. During
the powerful flares the fast processes cover huge areas which are
continued from the photosphere to the corona. Flares reflect the
energetics of magnetic fields and the behavior of parameters of
flares in the 11-year cycle are of considerable interest. We used
data from observations of $F_{10.7}$, SSN, and other solar indices
from the archives of NOAA National Geophysical Data. Analysis in the
EUV-region of the spectrum is the essential part of a studying of
the processes of solar flares. The radiation in this spectral region
(generated in the upper layers of the chromosphere and the
transition layer) characterizes the conditions of these regions of
the solar atmosphere during the development of flare. Observations
in the EUV-region are significantly complementing our knowledge on
the mechanisms of development of flares [Nusinov and Kazachevskaya,
2006].

Avakyan and Voronin, [2012] have analyzed the sun-weather-climate
connections. They showed that there is a link between the processes
occurring in the atmosphere of the earth and the level of
solar-geomagnetic activity including solar flares. Thus, the task of
studying of variations in solar activity is now becoming very
relevant in connection with the problem of global warming in which
the solar activity (in particular the flare activity) along with the
human activity factor plays a significant role.

The aim of the work is the analyzing of the flare activity of the
Sun in cycle 24 and the comparison of the fluxes during the
evolution of the large flare of X6.9 class according to GOES-15
(Geostationary Operational Environmental Satellite) and SDO (Solar
Dynamics Observatory) observations in the range of wavelengths from
the soft X-ray to the extreme ultraviolet.

\vskip12pt
\section{FLARE ACTIVITY IN CYCLE 24}
\vskip12pt

We have analyzed the data of the catalogue «Preliminary Current
Catalog of Solar Flare Events with X-ray Classes M1 -- X17.5 24-th
cycle of Solar Activity (I.2009 -- XII.2015)» available at
http://www.wdcb.ru/stp/data/
Solar\_Flare\_Events/Fl\_XXIV.pdf.
 We
have selected the 115 flares of M5 -- X7 classes and 45 flares of X1
-- X7 classes, which occurred on the Sun from February 2010 to
December 2015. We have also calculated the solar Flare Index (FI) in
accordance with the generally accepted standards from almost during
the whole 24-th cycle is approximately proportional to the total
energy emitted by a flare: $FI = i \cdot t$, where i is the
intensity scale of importance defined in [Atac, 1987] and t is the
duration of the flare in minutes. In Figure 1 the variations of the
SSN, the $F_{10.7}$  and the FI in the cycles 21 -- 24 are shown.
The general level of activity (according to observations of the SSN
and the $F_{10.7}$) in cycle 23 is slightly reduced, and in cycle 24
it decreases rather strongly (by about 30\% in the maximum of cycle
24 compared to the previous maximums).

\begin{figure}[tbh!]
\centerline{
\includegraphics[width=140mm]{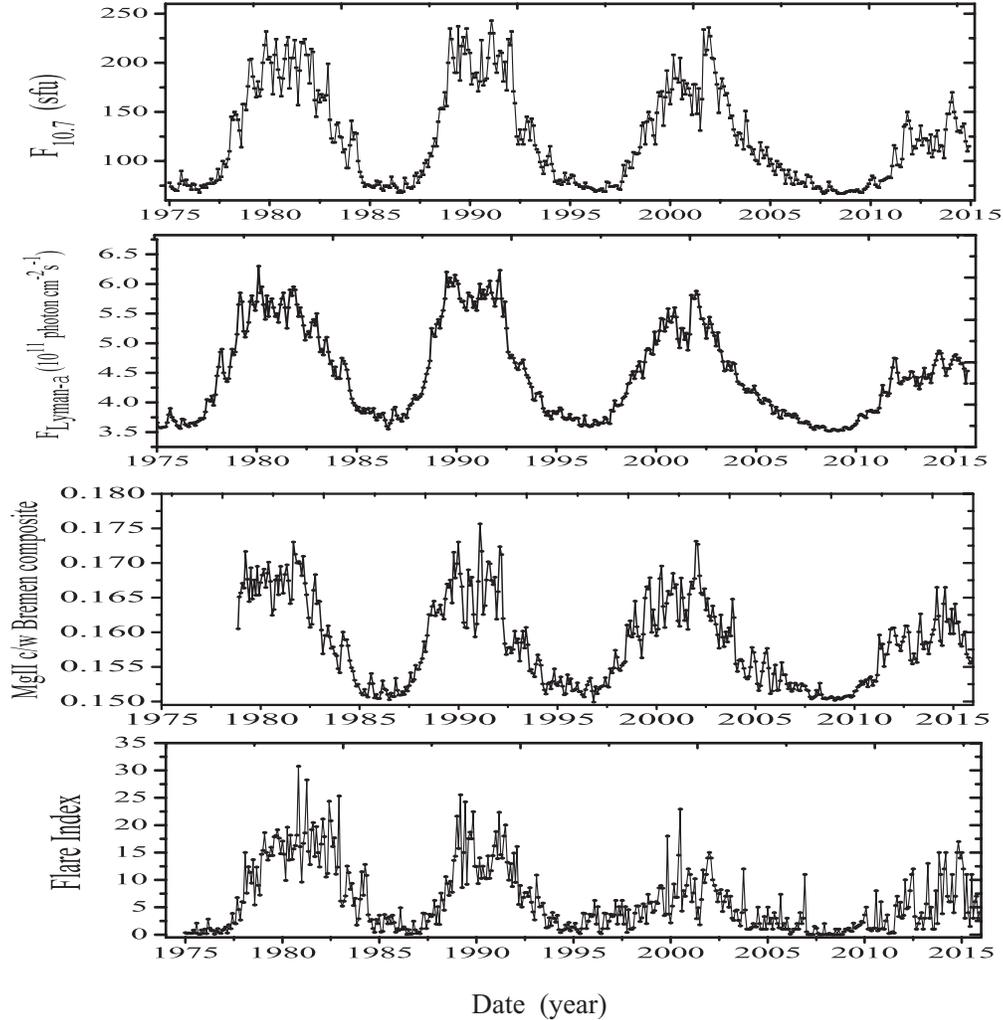}}
 \caption{Time series of solar activity indices -- monthly average values of SSN, $F_{10.7}$ and Flare Index (FI).}
{\label{Fi:Fig1}}
\end{figure}

In [Bruevich et al., 2014] it was shown that the dependence of FI
versus the overall level of solar activity in cycles 21 -- 23 varies
from cycle to cycle. Figure 1 shows that the FI is very much
decreases in cycle 24, its value in the maximum of cycle 24 is
almost 2 times less than in the maximums of  cycles 21 and 22 [NOAA,
2016].

Atac [1987] noted that FI is one of the best indicators of activity
variations in the chromosphere. It is of value as a measure of the
short lived (minutes to hours) activity on the Sun. This feature
makes the FI a suitable full disk solar index for comparison with
similar solar indices which reflect different physical conditions
from the different layers of the solar atmosphere.

\begin{figure}[tbh!]
\centerline{
\includegraphics[width=140mm]{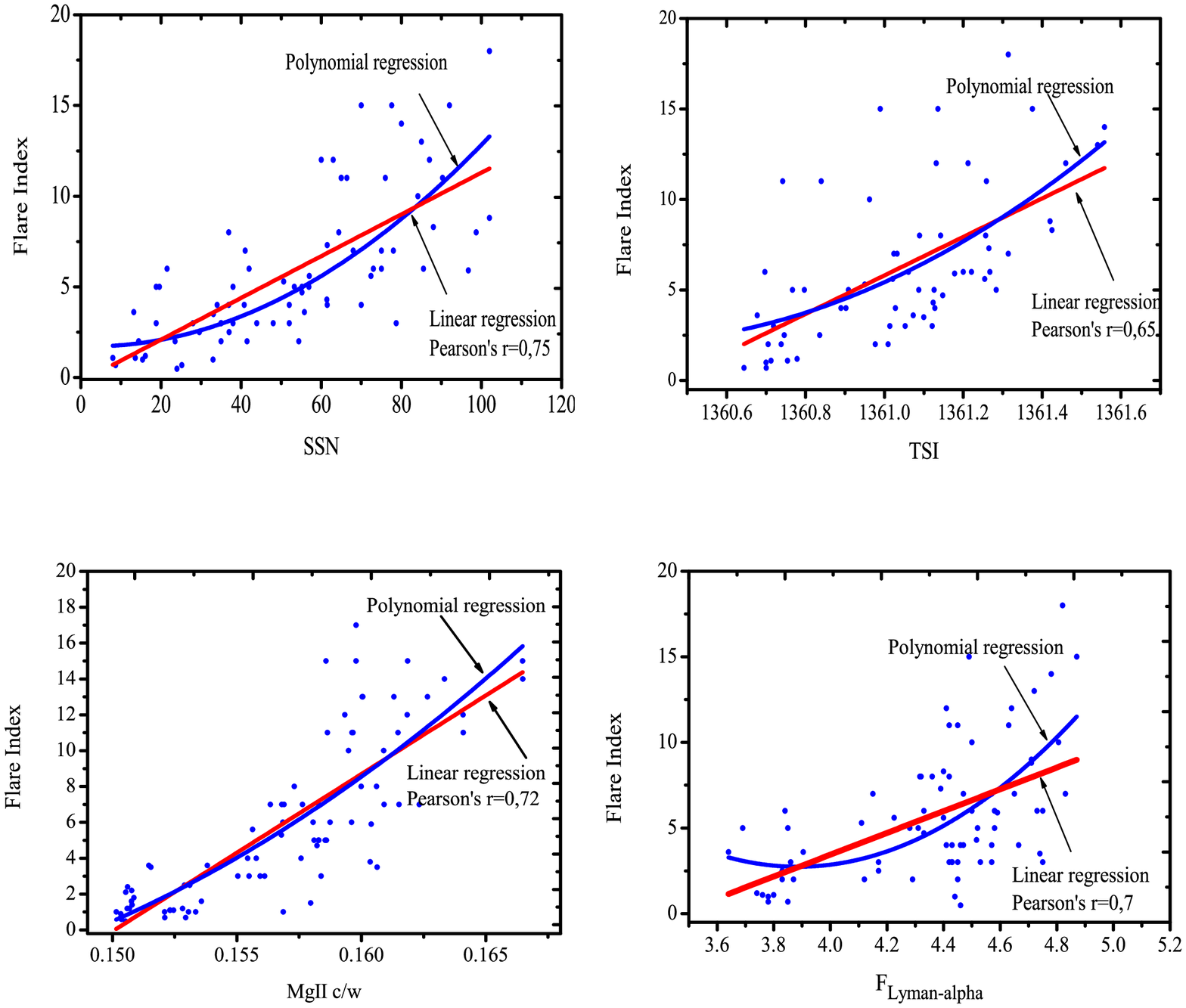}}
 \caption{The relationship of monthly values of FI versus SSN, TSI, MgII c/w and $F_{L \alpha}$  in cycle 24.
 The linear and polynomial regressions are shown.}
 {\label{Fi:Fig2}}
\end{figure}

Figure 2 shows the relationships between the FI versus the SSN, the
solar constant TSI, the Mg II c/w (280 nm) index and the radiation
flux $F_{L \alpha}$ . Good correlation was found between FI and SSN,
Mg II c/w index and the $F_{L \alpha}$, but not so good correlation
exists between FI and TSI. The coefficients of linear correlation of
FI versus other indices in cycle 24 are about 0.65 -- 0.75. A closer
correlation between FI and other solar indices were observed in
cycles 21 -- 23, the coefficients of linear correlation were about
0.75 -- 0.85, [Bruevich et al. 2014].

\begin{figure}[tbh!]
\centerline{
\includegraphics[width=140mm]{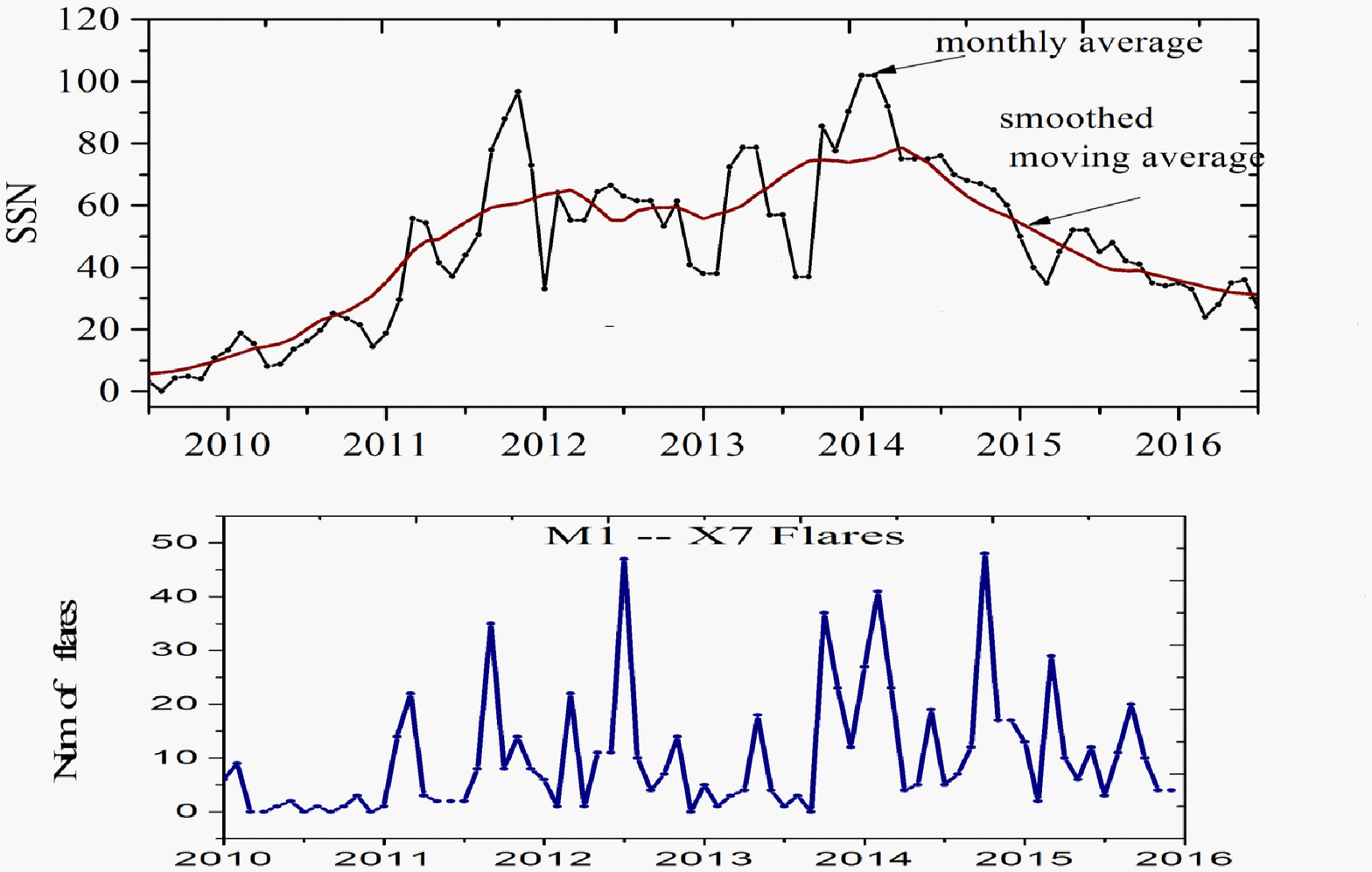}}
 \caption{The distribution of monthly average of the number of flares of M1--X7 classes (about 700 flares)
 in cycle 24. The maximum of the number of flares (M1--X7 classes) do not coincide with the maximum of the
  number of spots in cycle 24, with the exception of 2014.}
{\label{Fi:Fig}}
\end{figure}

Figure 3 shows the distribution of monthly average of number of
flares M1 -- X7 (more than 700 flares) in cycle 24. It is seen that
the maximum number of flares (M1 -- X7) does not coincide with the
maxima of the number of spots in cycle 24, with the exception of
2014. We noted that the greatest number of flares in cycle 24
occurred in the decline phase after the first maximum, during the
second maximum and in the decline phase after the second maximum.
The largest flares of cycle 24 of the X-ray class > X2.7 occurred on
rise phase and in the maximum. But in cycles 21 -- 23 the largest
flares of the X-ray class > X9 were observed on the decline phases
of cycles 21 and 23, as well as at the maximum of cycle 22,
[Klystov, 2014].

\begin{figure}[tbh!]
\centerline{
\includegraphics[width=140mm]{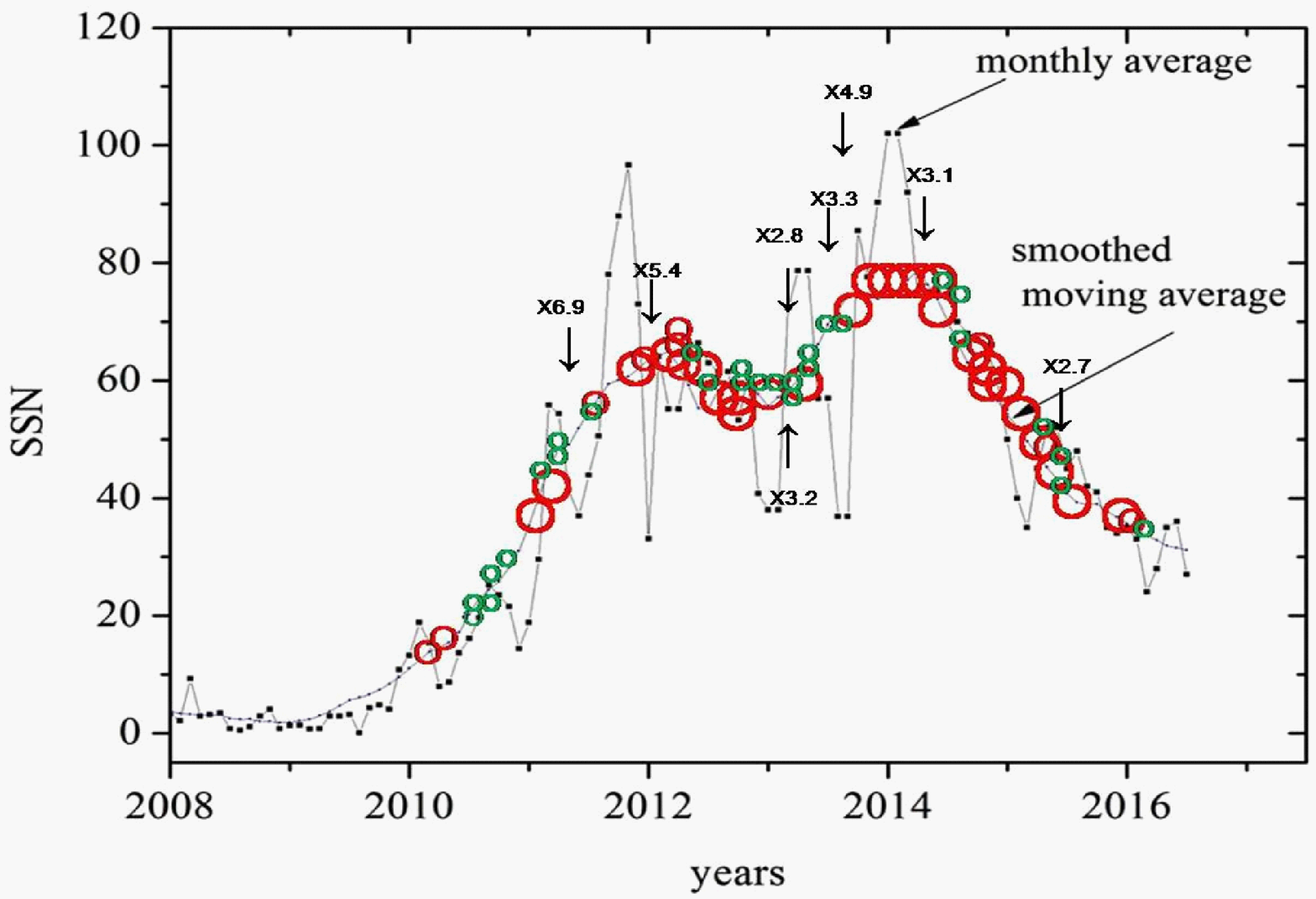}}
 \caption{The distribution of the number of flares (monthly average) in cycle 24.
 The arrows point out to the strongest flares. Small circles shows the months when numbers of flares
 (M1 -- X7 classes) for a month are less than 10. The large circles shows the months when numbers
 of flares are more than 10 flares per month.}
{\label{Fi:Fig}}
\end{figure}

The distribution of the number of flares (monthly average) in cycle
24 is shown in Figure 4. The strong flares (class > X2.8) are
pointed with the arrows. The months when numbers of flares (M1 -- X7
classes) for a month are less than 10 are pointed with the small
circles. The months when numbers of flares are more than 10 flares
per month are pointed with the big circles. Note then the greatest
number of flares occurred after the first maximum, in the second
maximum, and at the decline phase of cycle 24. The largest X-ray
flares of classes X2.8 and more strong flares occurred during the
rise phase and in the maximum phase of cycle 24.

There is very important to the study of such parameters of the
flares. Their integrated (in time) energy and their duration of
energy release that carry a lot of information about the energy of
flares over the entire range of energy in addition to the
conventional characteristics of flares, characterizing their class
in the X-ray band. X-ray observations of solar flares is more
suitable for statistical analysis of their energetics. They have the
advantage over the optical solar observations of flares, where it is
necessary to integrate over the surface of the flares.

\begin{figure}[tbh!]
\centerline{
\includegraphics[width=140mm]{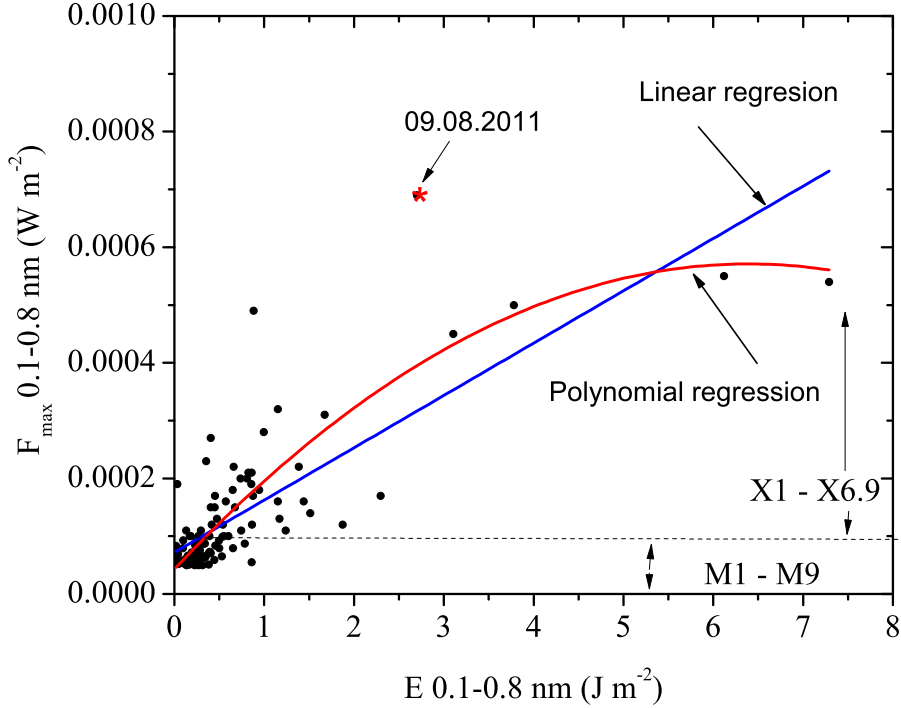}}
 \caption{The dependence of the flare's amplitude $F_{max}$ versus the total energy of the flare E
 according GOES-15 0.1 -- 0.8 nm observational data. The analysis is presented for the flares of M5 -- X7 classes.
 The flare 09.08.2011 is pointed with the asteric.}
{\label{Fi:Fig}}
\end{figure}

In Figure 5 we can see that flare energy E, as the most important
feature of a flare impact on the earth's atmosphere, has not the
close relationship with the amplitude of the flare in soft X-rays.
We have calculated the values of the total energy $E (J \cdot
m^{-2})$ in the range 0.1 -- 0.8 nm at the level of the earth's
atmosphere for 115 flares. Evaluation of energy of optical flares in
[Hudson, 1991] showed that the distribution of their integral in
time of the values of the energies can be represented by a power
function $N \sim E^{-\beta}$. The universality of this distribution
was proved for X-rays flares. The measure of the energy spectrum
$\beta$ was markedly changed in different phases of the 11-yr cycle.
In [Sotnikova, 2010] the study of energy spectra of flares in the
soft X-rays band during solar cycles 21 and 22, for each year
separately, was made. It was shown that the spectrum index $ \beta =
dlogE/dlogN$ depends on the rise phase, the maximum and the minimum
of the 11-yr cycle. It was also obtained that the yearly average
energy of flares for each year of the cycle significantly vary with
the phase of the cycle, increasing in overall from minimum to
maximum of the cycle.

\begin{figure}[tbh!]
\centerline{
\includegraphics[width=140mm]{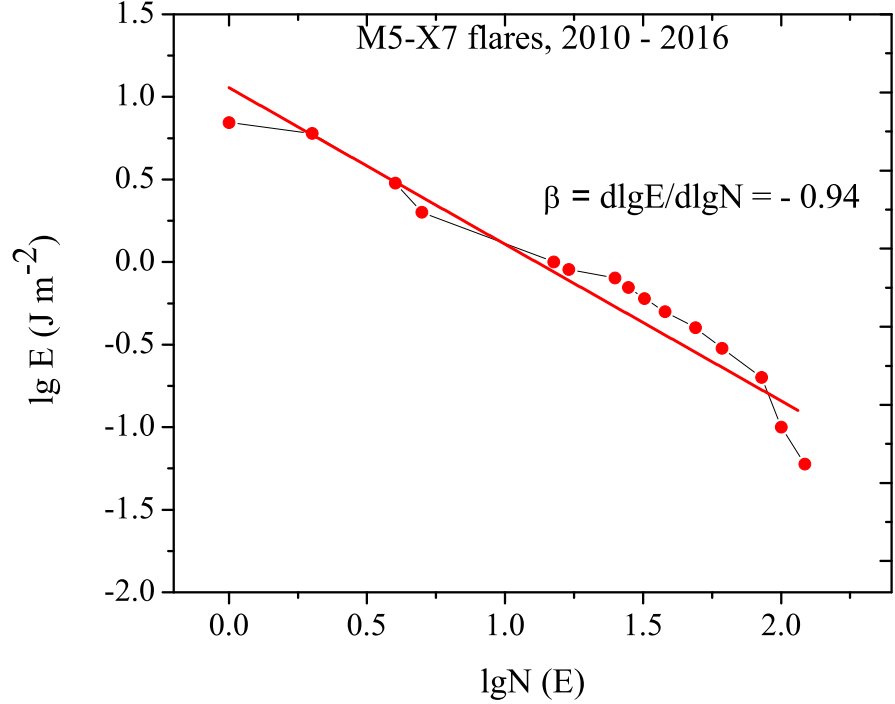}}
 \caption{The energy spectrum of the flares is the number of flares with a certain total energy
 (in the soft X-ray interval 0.1 -- 0.8 nm) versus the value of this total energy, constructed by us using
 data from the satellite GOES-15. The distribution of the number of flares with a certain energy depending
 on the energy values of the flares, we can imagine as the exponential function $N \sim E^{-\beta}$.}
{\label{Fi:Fig}}
\end{figure}

For 115 flares of M5 -- X7 class of cycle 24 we have built the
energy spectrum. The energy spectrum of the flares -- is the number
of flares with a certain total energy (in the soft X-ray interval
0.1 -- 0.8 nm) versus the value of this total energy -- has been
constructed by us using data from the satellite GOES-15. Figure 6
shows that the absolute value of $\beta$ in cycle 24 corresponds to
the phase of maximum of the 11-yr cycle according to [Sotnikova,
2010]. The straight line marks the dependence $\beta = dlogE/dlogN =
- 0.94$.

\vskip12pt
\section{THE LARGEST FLARE OF X-RAY CLASS X6.9 IN CYCLE 24 }
\vskip12pt

Schrijver [2016] has studied large flares (M5 -- X7 classes) based
on SDO/AIA observations and showed that full-disk emission images of
the corona and transition region, the Measure of Emission (ME) of
lines, the dynamics of the surface magnetic fields differed in
different flares and out of flares. The AIA (Atmospheric Imaging
Assembly) was launched as a part of NASA Solar Dynamics Observatory
(SDO) mission; irradiance products from AIA are available at AIA
website, https://sdo.gsfc.nasa.gov/data/aiahmi/. AIA will advance
our understanding of the mechanisms of solar variability and of how
the energy of the Sun is stored and released into the heliosphere
and geospace, [Lemen et al., 2012].

The EVE (Extreme ultraviolet Variability Experiment) instrument, and
its calibration, are well understood [Woods et al., 2012]. A key
aspect of the SDO satellite in geosynchronous orbit (GEO) is that it
provides continuous solar observations with very few Earth eclipses.
Irradiance products from EVE are available in a variety of
wavelength resolutions and time cadences from the EVE website,
http://lasp.colorado.edu/eve/data\_access/service/

For large flares, a significant amount of heating energy could be
provided during the decay phase, sometimes more than during the rise
phase observing the movement of X-ray emission source, put forward
that the continuous heating into the decay phase is in the form of
"sequential heating of new loops in the flare region" [Qiu and
Longcope, 2016]. We have analyzed the flare of X-ray class X6.9 to
compare the fluxes and the development of the flare in the range of
wavelengths from the soft X-rays ($\lambda$ = 0.1 -- 0.8 nm) until
to the extreme ultraviolet ($\lambda \leqslant 30.4$ nm), see Figure
7 and Table 1. The EUV solar irradiance is made by the EVE
instrument aboard NASA's Solar Dynamics Observatory.

On the example of X6.9 flare (09.08.2011), we analyzed the sequence
of the beginning of the impulsive phase of flare in different lines
of the EUV-range. The excellent SDO observations with a resolution
of 1 minute allow to determine the delay between the beginning of
the flare  in different lines.

\begin{figure}[tbh!]
\centerline{
\includegraphics[width=140mm]{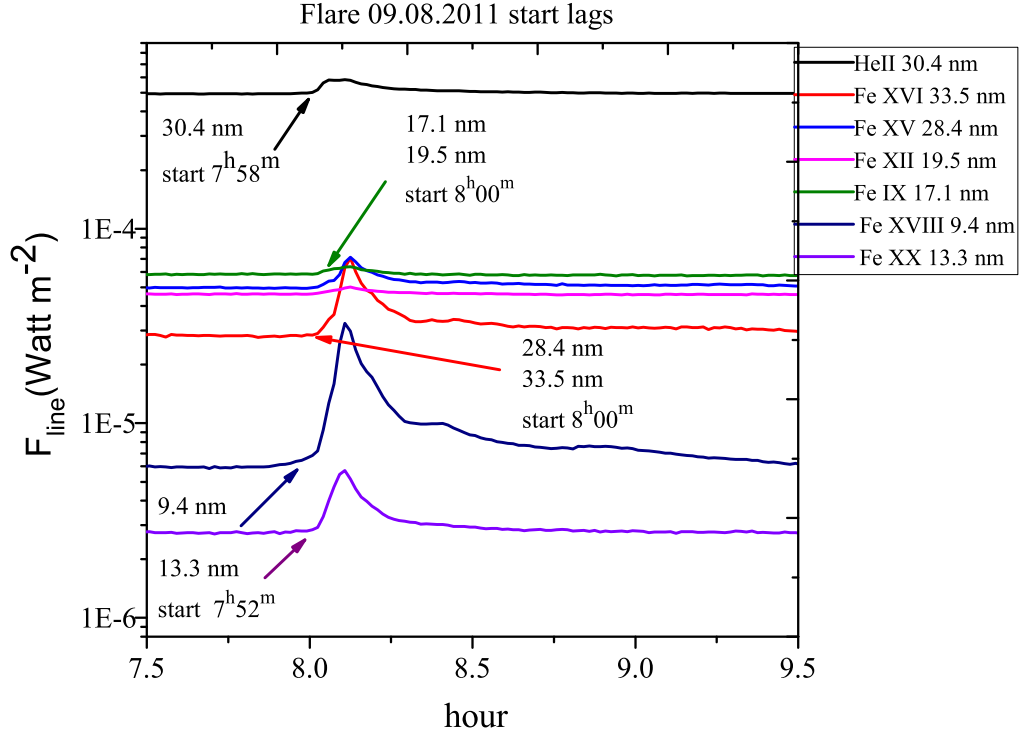}}
 \caption{The development of the largest flare in cycle 24 of X6.9
 class observed 9.08.2011 at different wave-lengths of EUV-range. Observations of SDO/EVE.}
{\label{Fi:Fig}}
\end{figure}

Figure 7 shows that the level of the absolute values of fluxes in
the studied lines of the EUV-range according to the SDO/EVE data
varies from  $ 3 \cdot 10^{-5} (Watt\cdot m^{-2}$) at the line 33.5
nm to $ 5 \cdot 10^{-4} (Watt\cdot m^{-2}$) from the strongest line
of the EUV-range 30.4 nm. In Figure 7 we can see that the beginning
of the studied flare X6.9 in different lines occurs at different
times, and the time of maximum in these lines occurs almost
simultaneously. For this flare according to TIMED (Thermosphere,
Ionosphere, Mesosphere, Energetics and Dynamics) observations
(http://science.nasa.gov/missions/timed/) the increase of the
radiation flux was: in the range 0.1 -- 7 nm -- 44 000 times, in the
line 30.4 nm (HeII) $\sim 20\%$; in the line 121.6 nm ($Lyman -
\alpha) \sim 8\%$.

Table 1 shows the summary of SDO/EVE observations in 8 EUV-lines and
in the X-ray interval of 0.1 -- 0.8 nm in which we can see the lags
in different lines of flare beginning. We note that the moments of
beginning of flare in different lines of the EUV-range (SDO/EVE) and
in Soft X-ray 0.1 -- 0.8 nm (GOES-15) lagged each other in different
lines at 2 -- 5 minutes.

We can see from the Table 1 that this X6.9 flare began, first in
lines Fe XX (13.3 nm) and in line Fe XVIII (9.4 nm). These lines are
formed in flaring corona at a temperature of $log(T) = 6.8 \div
7.0$. Then the flare starts in the range of 0.1-0.8 nm. In the lines
Fe XVI (33.5 nm), Fe XV (28.4 nm), Fe IX (17.1 nm) and in the line
He II (30.4 nm) the flare starts later. These lines are formed in
active-region corona and transition region at a temperature of
$log(T) = 5.0 \div 6.0$, see Table 2, Fig. 8.

Table 2 shows how the different ions which are  involved in the
formation of bright EUV-spectral lines are distributed with height
in solar atmosphere. Figure 8 in accordance with Table 2
demonstrates the distribution with height of the regions of the
EUV-lines formation according to the Atmospheric Imaging Assembly
(AIA) provides multiple simultaneous high-resolution full-disk
emission images of the corona and transition region, in the EUV-band
lines: FeXVIII (9.4 nm), FeVIII, XXI (13.1 nm), FeIX (17.1 nm),
FeXII, XXIV (19.3 nm), FeXIV (21.1 nm), HeII (30.4 nm), and FeXVI
(33.5 nm).

\begin{table}
\caption{The largest flare of cycle 24 -- X6.9 according to the X-ray classification.
The moments of flare beginning and flare maximum in different lines of the EUV-range according
to SDO/EVE and in soft X-ray (SXR) 0.1--0.8 nm according to GOES-15.}
\vskip12pt
\begin{tabular}{clclclcl}

\hline

  Wavelength (nm) &    Ion, Satellite  &    Flare begin. (UT)  &     Flare max. (UT)   \\

\hline

9.4  & FeXVIII, SDO   & 7:50-7:51 & 8:05-8:06  \\
13.3 & FeXX, SDO      & 7:52-7:53 & 8:06-8:07  \\
 0.1--0.8 &    SXR, GOES-15  &  7:54-7:55    &   8:05-8:06   \\
13.1  &    FeVIII, SDO &  7:56-7:57    &  8:06-8:07   \\
30.4  &    HeII, SDO &  7:58-7:59   &  8:05-8:06  \\
33.5  &    FeXVI, SDO    & 8:00-8:01   &  8:06-8:07  \\
28.4  &   FeXV, SDO  & 8:00-8:01   &  8:07-8:08  \\
19.5  &   FeXII, SDO         & 8:00-8:01   &  8:06-8:07  \\
17.1  &   FeIX, SDO      & 8:00-8:01   &  8:07-8:08 \\
\hline

\end{tabular}
\end{table}

\begin{table}
\caption{The primary ions observed by SDO. Many are species of iron covering more than a decade in coronal temperatures.}
\vskip12pt
\begin{tabular}{clclclcl}

\hline

 Channel (nm) &    Primary ion(s)  &   Region of atmosphere  &     log(T)   \\

\hline

30.4 & HeII   & transition region, upper photospher & 4.7  \\
160.0 & CIV + cont.   & quiet corona, upper photospher& 5.0 \\
17.1 & FeIX  & quiet corona, transition region  & 5.8 \\
19.3 &    FeXII, XXIV  & corona and hot flare plasma   &  6.2, 7.3   \\
33.5 &      FeXVI & active-region corona   &    6.4   \\
9.4&    FeXVIII &  flaring corona   &  6.8\\
13.1  & FeVII, FeXX &  transition region,flaring corona   & 5.6, 7.0  \\
\hline

\end{tabular}
\end{table}

\begin{figure}[tbh!]
\centerline{
\includegraphics[width=140mm]{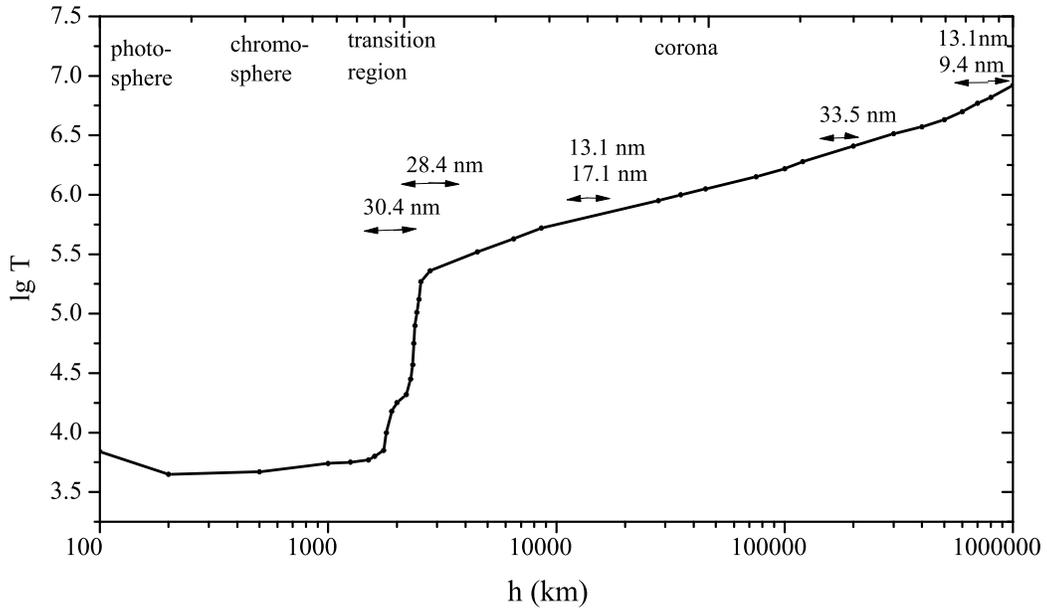}}
 \caption{The temperature of solar atmosphere change of with height.
 The regions of the solar atmosphere corresponding to the maximums of fluxes in the lines HeII 30.4 nm,
 FeXV 28.4 nm, FeXXI 13.1 nm, FeIX 17.1 nm, FeXVI 33.5 nm and FeXVIII 9.4 nm are marked.}

{\label{Fi:Fig}}
\end{figure}

Figure 7 shows that there are the shifts of the moment of the flare
beginning in different spectral lines according SDO/EVE and GOES-15
observations of the flare X6.9 that occurred 09.08.2011. The same
flare was studied in detail in [Sharykin, et al., 2015], were the
Plasma Heating to Ultrahigh Temperatures was analyzed. A model of
flaring region in the two-temperature approximation is presented in
[Sharykin, et al., 2015]. It was assumed that in the upper part of
the flare loop exists the area with the super hot (sh) plasma. The
observed plasma in the sh-region is assumed as a source of
non-thermal electrons and is associated with their acceleration. The
sh-region lies closest to the proposed site of the initial energy
release in the corona (the region of magnetic reconnection). The
population of nonthermal electrons is entered to lower levels till
hot-region and further into the dense part of the solar atmosphere.
This leads to evaporation of the chromosphere. The transfer
processes occur along the force lines of the magnetic field, the
heat transfer occurs inside the flare loop. The fluxes in coronal
lines increase during the process of plasma heating in the flare due
to the increase of the emission measure. This model corresponds to
our study, according to which the fluxes increase, first in lines
9.4 nm and 13.1 nm and in the end when nonthermal electrons reach
the chromospheric condensation, the fluxes increase in lines 28.4 nm
and 30.4 nm. Table 1 shows that this flare first starts brightening
in the lines formed in the corona at a temperature of $log(T) = 6.8
\div 7.0$ and then starts in the lines formed in the low corona and
transition zone at a temperature of $log(T) = 5.0 \div 5.5$.

\bigskip

\vskip12pt
\section{CONCLUSIONS}
\vskip12pt

The flare activity in cycle 24 was significantly lower than in
cycles 21 -- 23 according to the observational data (GOES-15, M1 --
X7 class) and SDO/EVE (EUV -- range). Our analyzing of the solar
activity in cycle 24 and the study of the large flare of X6.9 class
in the range of wavelengths from the soft X-ray until to the extreme
ultraviolet showed the following:

-- The peculiarity of the investigated flare activity is that the
coefficient of linear correlation k for the dependences "FI versus
other solar indices" calculated for the 2010 -- 2016 is equal $k
~\sim 0.65 \div 0.75$ in cycle 24, less than k in cycles 21 -- 23
($k \sim 0.75 \div  0.85$), see [Bruevich et al., 2014].

-- Most of large flares (X-ray class >X2.7) occurred in the rise
phase and in the maximum of cycle 24. Unlike the large flares in
cycles 21, 22 and 23, where the greatest number of flares were
observed during the decline phases.

-- It is shown that for the analysis of the impact of solar flares
on the atmosphere of the  earth  it is necessary to consider the
total energy of flares -- E. Unlike the $F_{max}$ -- the maximum
amplitude of the flare flux in the range of 0.1 -- 0.8 nm, see
Figure 5.

-- We have constructed the energy spectrum of flares in cycle 24
using the data from the satellite GOES-15 for the 115 flares of the
X-ray class > M5. This index of the energy spectrum $\beta$ for 115
flares in cycle 24 corresponds to the phase of maximum of the cycle.

-- The SDO/EVE and GOES-15 observations for the largest flare (X6.9)
of cycle 24 that occurred 09.08.2011 showed that the  shift of the
moment of the flares beginning in the spectral lines 9.4 nm and 13.3
nm was for about 3 -- 4 min relative to the moment of the flares
beginning in 0.1 -- 0.8 nm region, was about 5 -- 6 minutes relative
to the start time of the flare in the spectral line 30.4 nm and was
about 7 -- 8 min relative to the start time of the flare in the
spectral lines of 17.1 nm, 19.5 nm, 28.4 nm and 33.5 nm, see Table
1.

\end{document}